\documentclass[pra,twocolumn,showpacs,preprintnumbers,amsmath,amssymb,tightenlines,epsfig]{revtex4}
\usepackage[utf8]{inputenc}
\usepackage{graphicx}
\usepackage{color}
\usepackage{xcolor}
\usepackage{physics}
\usepackage[percent]{overpic}
\usepackage{ragged2e}
\usepackage{amsmath}
\usepackage{mathrsfs}

\newcommand{\rvec}[1]{{\mathbf{r}}}

\newcommand{\expec}[1]{\langle #1 \rangle}

\newcommand{\sub}[2]{{#1}_{\mbox{\!\! \scriptsize #2}}}

\def\beq{\begin{equation}}
\def\eeq{\end{equation}}

\def\CR{\nonumber\\[0.15cm]}

\newcommand{\rref}[1]{Ref.~\cite{#1}}
\newcommand{\fref}[1]{Fig.~\ref{#1}}
\newcommand{\frefp}[2]{Fig.~\ref{#1}~(#2)}

\newcommand{\eref}[1]{Eq.~(\ref{#1})}

\newcommand{\sref}[1]{section~\ref{#1}}

\newcommand{\cref}[1]{chapter~\ref{#1}}

\newcommand{\Cref}[1]{Chapter~\ref{#1}}

\newcommand{\aref}[1]{appendix~\ref{#1}}
\newcommand{\bref}[1]{(\ref{#1})}

\usepackage{ulem}  
\begin{document}

\title{Density correlations from analogue Hawking radiation in the presence of atom losses}
\author{Yash Palan}
\affiliation{Department of Physics, Indian Institute of Science Education and Research, Bhopal, Madhya Pradesh 462 066, India}
\author{Sebastian~W\"uster}
\affiliation{Department of Physics, Indian Institute of Science Education and Research, Bhopal, Madhya Pradesh 462 066, India}
\email{sebastian@iiserb.ac.in}
\begin{abstract}
The sonic analogue of Hawking radiation can now be experimentally recreated in Bose-Einstein Condensates that contain an acoustic black hole.
In these experiments the signal strength and analogue Hawking temperature increase for denser condensates, which however also suffer increased atom losses from inelastic collisions. To determine how these affect analogue Hawking radiation, we numerically simulate creation of the latter in a Bose-Einstein Condensate in the presence of atomic losses. In particular we explore modifications of density-density correlations through which the radiation has been analyzed so far. We find that losses increase the contrast of the correlation signal, which we attribute to heating that in turn leads to a component of stimulated radiation in addition to the spontaneous one. Another indirect consequence is the modification of the white hole instability pattern.
\end{abstract}

\maketitle
\section{Introduction} 
%
Hawking radiation \cite{hawking:hr1,hawking:hr2,visser:review} is a prominent prediction of quantum field theory in curved space time \cite{parker:QFT_in_curved_spacetime,birrell:Quantum_fields_in_curved_space}. The difficulties with observing the radiation from an astrophysical black-hole have been a key motivation for the development of the analogue gravity program \cite{barcelo:diffmetric, garay:prl, visser:review}. The latter is founded on the mathematical correspondence between sound propagation in a fluid medium and the propagation of quantum fields in curved spacetime \cite{unruh:bholes}. 

Applying that idea to gaseous Bose-Einstein condensates (BECs) as a quantum fluid \cite{garay:prl,barcelo:diffmetric,visser:analogue,barcelo:cpc}, analogue Hawking radiation (AHR) has now been observed by measuring density-density correlations to very high precision \cite{steinhauer:bhlaser,steinhauer:entanglement,steinhauer:thermal}. Exploiting these correlations as experimental signature
\cite{fabbri:corrfct} offers several advantages, such as a clear connection between the Hawking particle (phonon) and its partner \cite{fabbri:corrfct}, a link to entanglement \cite{steinhauer:entanglement} and the ability to discriminate condensate heating that could mask the thermal nature of AHR in temperature based measurements \cite{wuester:horizon,wuester:phonon_vs_AHR}.

However, also when observing AHR through correlations, signals are stronger when the surface gravity of the sonic black hole is larger \cite{fabbri:corrfct}. At fixed Mach number profile, the surface gravity increases for denser condensates. However, these are also subject to stronger atom losses \cite{wuester:phonon_vs_AHR}, most notably three-body losses \cite{adhikari:three_body_loss,Ueda:three_body_loss,hiroki:three_body_loss}, te rates for which scale cubic with density. Losses drive the quantum many-body state of the Bose gas away from its ground-state and thus also cause quasi particle creation \cite{dziarmaga:lossheating}, which could interfere with Hawking signals.

In this article, we explore how one-, two- and three-body losses affect the correlation signature of AHR. We find that the characteristic features that link sonic Hawking radiation to the black hole horizon persist also in the presence of losses. For this we utilize the truncated Wigner approximation \cite{steel:wigner,Sinatra2001,castin:validity,wuester:nova2,wuester:kerr,wuester:collsoll,ashton:loss,carusotto:Quantum} for the dynamics of fluctuations around the mean field of a BEC, which has been successfully applied earlier in the context of analogue gravity \cite{mayoral:white_hole, fabbri:wigner, visser:review,Tettamanti:study, deNova:bhlaser_instability}.
We find that correlation features are strengthened in simulations that include losses, which we attribute to an additional stimulated Hawking radiation component \cite{weinfurtner:stimulated_HR} due to loss-induced condensate heating \cite{dziarmaga:lossheating,wuester:phonon_vs_AHR}. Additional modifications of experimental observables by the losses are a change in the slope of the AHR tongue and the emergence of additional tongues and patterns due to instabilities at the white hole that are accelerated by the noise. 
These results show that the subtle interplay of multiple aspects of BEC quantum field dynamics is manifest in correlation patterns, and a careful comparison of numerical simulations and experimental results can thus provide insight also into features that are not directly pertaining to analogue gravity.

This article is organized as follows: a brief description of the sonic black hole scenario and the truncated Wigner method is provided in \sref{sec:TWA_sim_of_bh}. In \sref{sec:Density_correlations} we review the correlation observable that we focus on and the most important features it exhibits. Subsections therein describe the modification of these features due to atom losses, with strengthening of correlations in \sref{strong_correl_loss}, discussion of the slope of Hawking tongues in \sref{slope_change} and the white hole correlation pattern in \sref{whitehole_pattern}. Details regarding the truncated Wigner method have been summarized in \aref{TWAlosses} and \ref{TWA_correlations}, while details regarding white hole damping can be found in \aref{WH_damping}.

\section{Truncated Wigner simulation of sonic black hole} \label{sec:TWA_sim_of_bh}

\begin{figure}[htb]
   \begin{overpic}[width=\columnwidth]{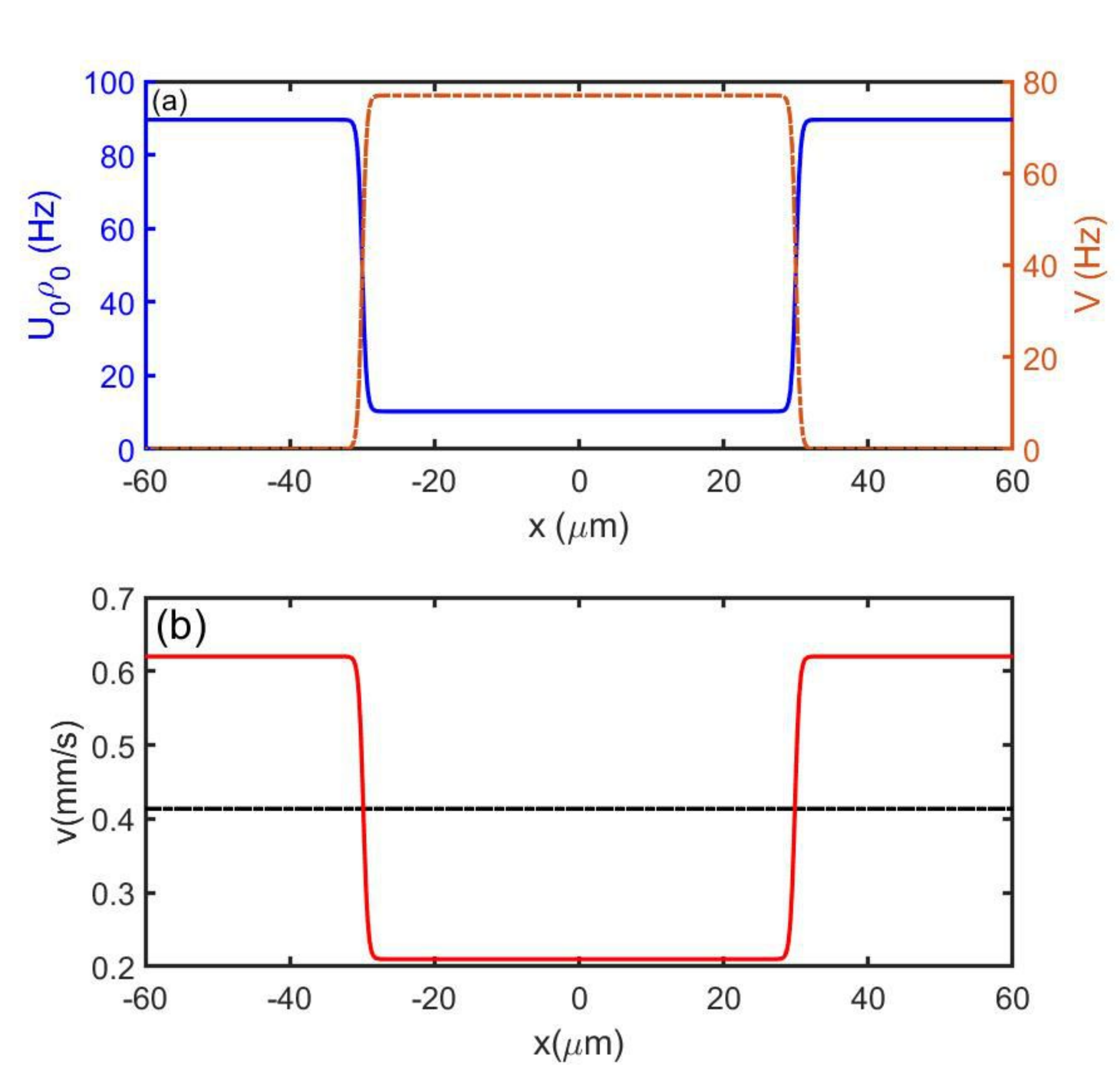}
    \put (23,26) {$v_0$}
    \put (23,40) {\color{red}$\sub{c}{sub}$}
    \put (50,13) {\color{red}$\sub{c}{sup}$}
     \put (23,60) {\color{orange}$\sub{V}{sub}$}
    \put (50,85) {\color{orange}$\sub{V}{sup}$}
     \put (18,83) {\color{blue}$\sub{U}{sub}\rho_0$}
    \put (50,62) {\color{blue}$\sub{U}{sup}\rho_0$}
\end{overpic}
    \caption{Potential, interaction and flow profiles of a BEC in a periodic 1D domain (on a ring). The ring is divided into two regions by the black hole horizon, at $x_{h}=-30$ $\mu$m and the white hole horizon, at $x_{w}=30$ $\mu$m. (a) Shape of the external potential $V(x)$ in \eref{eq:Vprofile} (orange dashed line) and interaction strength $U_0(x)$ in \eref{eq:Uprofile} (blue line) at times $t>0$ with $V_{\text{sub}}=0$, $U_{\text{sub}}= c^2_{sub}m/\rho_0$, $U_{\text{sub}}= c^2_{sup}m/\rho_0$ indicated as text. $V_{\text{sup}}$ is calculated using \eref{eq:condition_on_V_and_U}. The width of the step is $\sigma_{sp}=0.6$ $\mu$m. (b) Shows the spatial variation of the speed of sound $c_s(x)$ (solid red line), with $\sub{c}{sub}=0.62$ mm/s and $\sub{c}{sup}=0.21$ mm/s, and velocity of the condensate $v_0$ (black dashed line) for $t>0$.}
       \label{fig:GPE_simulation}
\end{figure}
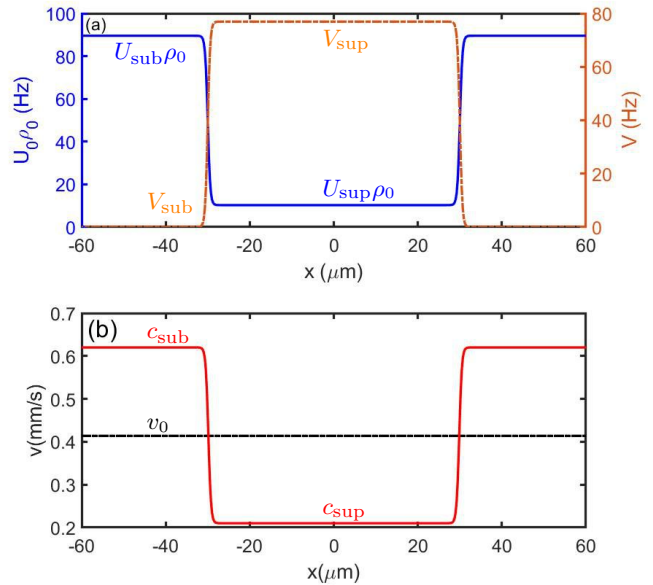

We consider a BEC of \textsuperscript{87}Rb atoms in a one-dimensional ring trap \cite{kunal:toroidal_trap_exp,mathey:toroidal_trap_exp}. Following the approach of \rref{fabbri:wigner} to yield tractable numerical simulations, we assume that both the external potential $V(x,t)$ and the interaction strength $U_0(x,t)$ can be varied along the coordinate $x$ along the ring and in time $t$. For atoms of mass $m$ the Gross-Pitaevskii equation (GPE) \cite{book:pethik} that describes the dynamics of the mean field $\psi(x,t)$ is then
\begin{equation}\label{eq:GPE}
    i\hbar\pdv{\psi}{t} = \left[-\frac{\hbar^2}{2m}\pdv[2]{x} + V(x,t) + U_{0}(x,t) |\psi|^2\right] \psi.
\end{equation}
For time $t<0$, we assume that the interaction strength is constant in space, $U_{0}(x)=\sub{U}{ini}$, and there is no external trapping potential, $V=0$. In this case, 
\begin{equation}\label{eq:background_solution}
	\psi(x,t<0) = \sqrt{\rho_0} e^{ik_0x}
\end{equation}
with density $\rho_0$ and condensate flow velocity $v_0=\hbar k_0/m$ related to the wave number $k_0$ is a solution of the time-independent GPE and thus a steady state of \eref{eq:GPE}. At $t=0$, we assume the interaction and external potential are modified (quenched) to 
\begin{equation}\label{eq:Vprofile}
	V(x,t>0)=\
	\begin{cases}
	    \sub{V}{sub} + \frac{(\sub{V}{sup}-\sub{V}{sub})\left[\tanh{\left(\frac{x-x_{h}}{\sigma_{sp}}\right)}+1\right]}{2}, x<0 \\
	    \sub{V}{sup} - \frac{(\sub{V}{sup}-\sub{V}{sub})\left[\tanh{\left(\frac{x-x_{w}}{\sigma_{sp}}\right)}+1\right]}{2}, x>0
	\end{cases}
\end{equation}
with $x_{h}$ the target location of the black hole horizon, $x_{w}$the white hole horizon and $\sigma_{sp}$ the length scale of the smoothened step function, shown also in \frefp{fig:GPE_simulation}{a}. Choosing further a constant combination
\begin{equation}\label{eq:condition_on_V_and_U}
	U_{\text{sub}}\rho_0 + V_{\text{sub}} = U_{\text{sup}}\rho_0 + V_{\text{sup}} \equiv C.
\end{equation}
we obtain the variation of the interaction strength $U_0(x,t>0)$ as 
\begin{align}\label{eq:Uprofile}
    U_0(x,t>0)&= \frac{C- V(x,t>0)}{\rho_0}.
\end{align}

This makes sure the chemical potential $\mu = \hbar^2k_0^2/(2m)+ V_{\text{sub}}+U_{\text{sub}}\rho_0 $ is constant and thus
 preserves \eref{eq:background_solution} as a solution of the time-independent GPE, albeit now an unstable one. This allows us to focus on the quench dynamics of quantum fluctuations around the mean field, without distractions by mean field dynamics.

The choice of potential divides the ring into a subsonic region, where $v_0 < c_{s}(x)=\sqrt{U_0(x)\rho_0/m}$, with speed of sound $c_s$, and a supersonic region where $v_0 > c_{s}(x)$. The transition from the subsonic to the supersonic region along the flow direction marks the black hole horizon, while the reverse marks the white hole horizon. 
Accordingly the interaction strength and external potential in \eref{eq:Vprofile} and \bref{eq:Uprofile} have been marked by subscripts $\{ \text{sub}, \text{sup} \}$, with "sub" referring to the subsonic region and "sup" referring to the supersonic region. The change of parameters described causes a sudden quench, from a flat analogue spacetime in a condensate without flow variation, to a spacetime containing a black-hole-white-hole pair, in a condensate with trans-sonic flow \cite{garay:pra,Finazzi_2010}.

Note, that while it is in principle realisable, the transition scheme from subsonic to supersonic flow discussed above has been chosen for numerical convenience only. To realize it, one would require a spatial dependence of the interaction strength by exploiting a Feshbach resonance with an inhomogeneous magnetic field, an accordingly tuned external potential $V(x)$, e.g.~optically, while working in a toroidal trap. It is experimentally much more straightforward to use a straight cigar shaped trap, in which the subsonic to supersonic transition occurs due to joint density and velocity variations induced by the external potential $V(x,t)$ only, keeping $U_0$ constant. This has hence been used in the actual experiment \cite{steinhauer:entanglement}. We expect all our results to pertain also to that scenario.

To numerically model analogue Hawking Radiation (AHR), we need to include quantum fluctuations of the condensate. This is done in the truncated Wigner Approximation (TWA) \cite{steel:wigner,Sinatra2001,castin:validity,blair:review}.
In the TWA method, the quantum state is represented by an ensemble of stochastic trajectories, with initial state given by 
\begin{equation} \label{eq:initial_state}
    	\psi(x,0) = e^{ik_0x} \left[ \sqrt{\rho_0} + \sum_{k \neq 0} \beta_k u_k e^{ikx} - \beta^*_k v_k e^{-ikx}  \right],
\end{equation}
where $\beta_k$ is a complex Gaussian random variable with $\overline{ \beta_k } = \overline{ \beta_k^2 } = \overline{ \beta_k \beta_{k'} }  = 0$ and $\overline{ \beta_k \beta^*_{k'} } =\delta_{kk'}[2\tanh{(\epsilon_k/2k_bT)}]^{-1}$, where $\overline{\cdots}$ denotes the stochastic average and $T$ is the temperature of the Bose gas. The Bogoliubov coefficients $u_k$ and $v_k$ are defined as usual in terms of the kinetic energy $E_k = \frac{\hbar^2k^2}{2m}$ and $\epsilon_k = \sqrt{E_k\left(E_k + 2U_0\rho_0\right)}$ according to $u_k \pm v_k = (E_k/\epsilon_k)^{\pm 1/2}$. 

The above stochastic initial state is then evolved using the TWA equation of motion, which follows from the masterequation for the system with the help of replacement rules \cite{steel:wigner}. Starting from the masterequation that includes atomic losses \cite{jack:loss}, following \cite{ashton:loss}, 
we discuss that procedure in \aref{TWAlosses}. The final result is
\begin{align}
      &d{\psi(x,t)} = dL_1 +  dL_2 +  dL_3  \label{eq:TWAeom} \\
      &-\frac{i}{\hbar}\left[-\frac{\hbar^2}{2m}\pdv[2]x +V(x,t) +U_0(x,t) |\psi(x,t)|^2\right] \psi(x,t) dt,  \nonumber
\end{align}
where decay and noise terms $dL_k$ for $k$-body loss are 
\begin{subequations}
\begin{align}
      dL_1 &= -\gamma_{1,1D} \psi dt+\sqrt{\gamma_{1,1D}}d\mathscr{W}, \label{eq:one_body_loss_term} \\
      dL_2 &= -\gamma_{2,1D} |\psi|^2 \psi dt +2\sqrt{\gamma_{2,1D}}|\psi| d\mathscr{W}, \label{eq:two_body_loss_term}\\
      dL_3 &= -\frac{\gamma_{3,1D}}{2} |\psi|^4 \psi dt + \frac{\sqrt{3\gamma_{3,1D}}}{\sqrt{2}} |\psi|^2 d\mathscr{W}. \label{eq:three_body_loss_term}
\end{align}
\end{subequations}
Here $\gamma_{1,1D}$, $\gamma_{2,1D}$ and $\gamma_{3,1D}$ are the effective one-body, two-body and three-body loss coefficients in 1D, respectively, see \aref{TWAlosses}. The symbol $d\mathscr{W}=d\mathscr{W}(x,t)$ denotes complex standard Wiener noise, with correlations $\overline{ d\mathscr{W}(x,t)} =0$, $\overline{  d\mathscr{W}(x,t)d\mathscr{W}(x',t') } =0$ and $\overline{  d\mathscr{W}(x,t)d\mathscr{W}^{*}(x',t') } =\delta(x-x')\delta(t-t')dt$. 

Finally, quantum field observables are extracted using symmetrically ordered averages \cite{steel:wigner}, such that for example the total atomic density is
\begin{align}
     \label{eq:density}
     \expec{\hat{\Psi}^\dagger(x)\hat{\Psi}(x)}&=\overline{|\psi(x)|^2} - \frac{1}{2}\delta_p(x,x),
\end{align}
where for a spatial domain  $-L\leq x <L$ the expression
\begin{equation} 
    \delta_p(x,x') = \frac{1}{L}\sum_{k} \left[ u^2_k e^{ik(x-x')} - v^2_k e^{-ik(x-x')}\right].
\end{equation}
 is a restricted basis commutator, discussed in \aref{TWA_correlations}. It has been shown in \rref{ashton:loss} that the truncation restricts the validity of the TWA method to scenarios where $|\psi(x)|^2 \gg \delta_p(x,x)$.
More details regarding the TWA method can be found in \cite{gardiner:Wigner_review}. It has been demonstrated first in \rref{fabbri:corrfct}, that the creation of analogue Hawking radiation can be modelled using the TWA.

\section{Density correlations} \label{sec:Density_correlations}
One of the most straightforward manifestations of AHR would be the re-heating of the condensate to the a analogue Hawking temperature 
\begin{equation}
    T_H=\frac{\hbar g_h}{2\pi k_b c_h}, \mbox{ with } c_h = c_s(x_h),
\end{equation} 
where $c_s$ is the speed of sound, $v_0$ is the velocity of the condensate, $x_h$ is the location of the black hole horizon, 
and $g_h$ the surface gravity of the sonic black hole. The latter can be found from \cite{fabbri:corrfct,Finazzi_2010}
\begin{equation}
    g_h = \frac{1}{2v_0} \left. \dv{[c^2(x)-v^2(x)]}{x} \right|_{x=x_h}.
\end{equation}
Demonstrating AHR thermally in this manner is however usually not practical, as the temperature is fundamentally limited by atomic loss processes \cite{wuester:horizon}
and remains less than the equilibrium temperature of loss induced heating \cite{wuester:phonon_vs_AHR}.

A popular observable that circumvents these problems is the density-density correlation function \cite{fabbri:corrfct} 
\begin{equation}\label{eq:G2def}
    G_2(x,x') = \frac{\langle \hat{\Psi}^{\dagger}(x) \hat{\Psi}^{\dagger}(x') \hat{\Psi}(x) \hat{\Psi}(x') \rangle}{\langle \hat{\Psi}^{\dagger}(x) \hat{\Psi}(x) \rangle  \langle \hat{\Psi}^{\dagger}(x') \hat{\Psi}(x')  \rangle}.
\end{equation} 
Density-density correlations $G_2(x,x')$ appear between a location $x$ outside the horizon and another one $x'$ inside the horizon since the Hawking particle 
and its anti-particle are created from the same entangling event at the event horizon. In contrast, pre-existing thermal excitations or those induced by losses are not expected to share any correlations that are linked to the horizon.
\begin{figure}[htb]
    \centering
    \includegraphics[width=0.99\columnwidth]{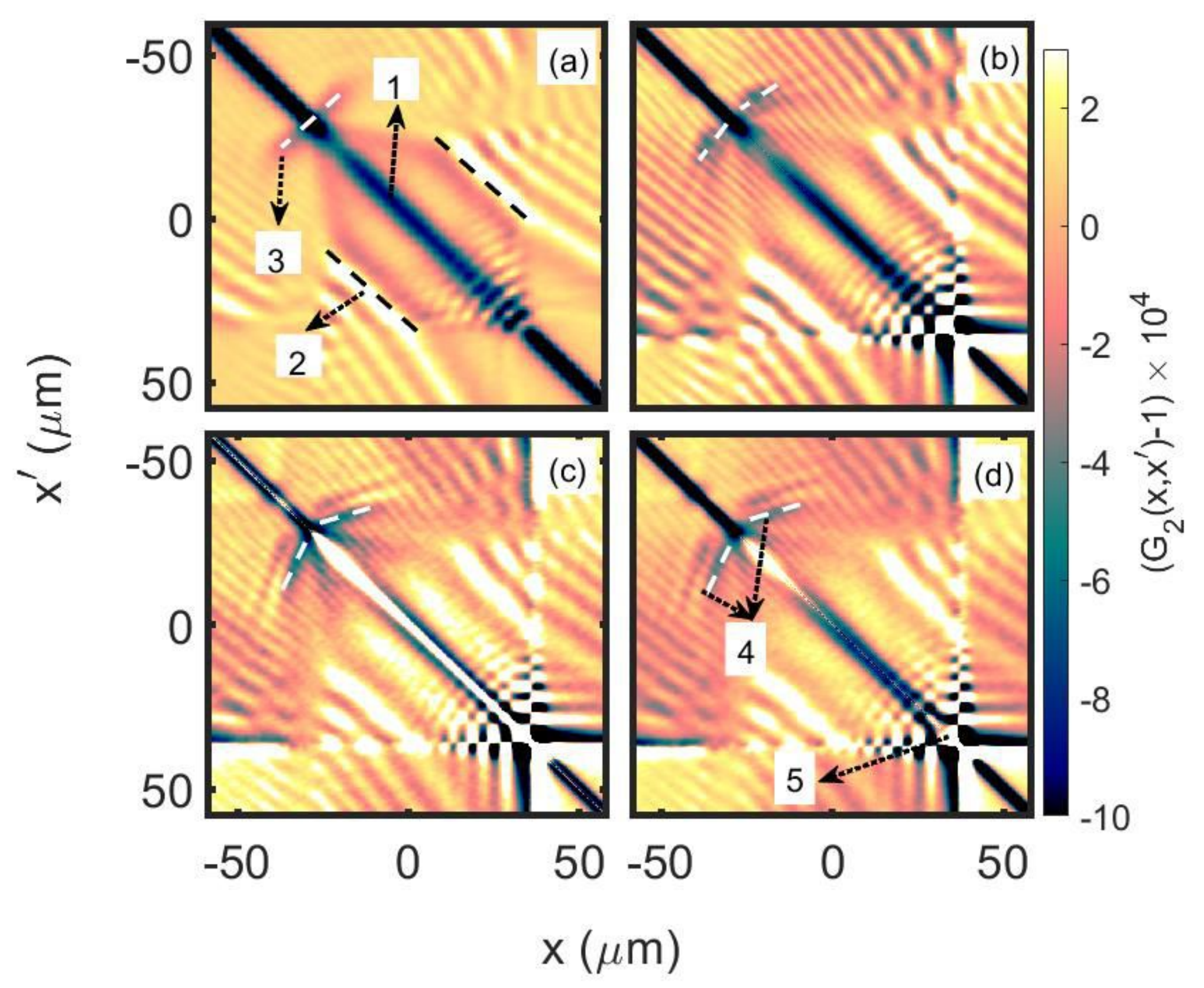}
    \caption{Density density correlations $G_2(x,x')-1$ near a sonic horizon (at $x=-30$ $\mu$m), including (a) no loss, (b) only one body loss, (c) only two body loss and (d) only three body loss. We show snapshots at $t=48$ ms after the initial quench. Movies for the same scenarios can be found in the supplemental material \cite{supp_material}. The features indicated by numbered arrows are discussed in the text.
    }
    \label{fig:atomic_losses}
\end{figure}

The experiments \cite{steinhauer:entanglement,steinhauer:thermal} thus relied on correlations \bref{eq:G2def} as a signature for AHR. 
The TWA method provides symmetrically ordered quantum correlations via averages of the stochastic wavefunction \cite{visser:review,steel:wigner}, which gives us the numerator of \eref{eq:G2def} as 
    \begin{align} \label{eq:G_2_correlation}
       &\langle\hat{\Psi}^{\dagger}(x)\hat{\Psi}^{\dagger}(x') \hat{\Psi}(x) \hat{\Psi}(x')\rangle =  \overline{|\psi(x)|^2|\psi(x')|^2}\CR
       & - \frac{1}{2}\Big[ \overline{ \psi^*(x) \psi(x')}  \delta_p(x,x') +  \overline{ \psi^*(x') \psi(x)} \delta_p(x',x)  \CR
       &\hspace{0.7cm}+\overline{ |\psi(x) |^2} \delta_p(x',x') + \overline{ |\psi(x') |^2} \delta_p(x,x) \Big]\CR 
       &+\frac{1}{4}\Big[\delta_p(x,x)\delta_p(x',x') + \delta_p(x,x') \delta_p(x',x) \Big].
    \end{align}
    %
The elements of the denominator can be calculated from \bref{eq:density}.

In this article, we compare the correlation signatures of AHR with and without the inclusion of atomic losses. These are shown in \fref{fig:atomic_losses}, using $\sub{N}{traj}=200000$ stochastic trajectories, i.e.~solutions of \eref{eq:TWAeom} at $t=0$. The condensate flow velocity is $v_0 = 0.415$ mm/s, with speed of sound in the subsonic and supersonic regions as $\sub{c}{sub} = 0.62$ mm/s and $\sub{c}{sup} = 0.21$ mm/s, respectively. The circumference of our ring or length of the 1D domain is chosen as $L=60$ $\mu$m, and the mean density prior to the quench at $t<0$ used in the simulations is $\rho_0 = 66.6$ $\mu\text{m}^{-1}$. Finally, the 3D loss coefficients were set to $\gamma_{1,3D}=3.096$ $\text{s}^{-1}$, $\gamma_{2,3D}=0.39$ $\mu\text{m}^3/\text{s}$, $\gamma_{3,3D}=0.06$ $\mu\text{m}^6/\text{s}$, as discussed in \rref{Pendse_solitondecoh_PhysRevA} and references therein. These were then converted to the effective 1D loss rates by using \eref{eq:reduced_loss_rates}, assuming a transverse trapping frequency of $130$ Hz. Solutions of \eref{eq:TWAeom} and averages \bref{eq:G2def} are obtained using the high level language XMDS \cite{xmds:paper}. To smoothen the correlations, they have been convolved with a Gaussian filter with kernel width $\approx1.7$ $\mu$m.  

Let us first describe the features in the correlation function $G_2(x,x')$ for the basic scenario without losses in \frefp{fig:atomic_losses}{a}, which have been observed before \cite{fabbri:wigner,carusotto:Quantum}:
\begin{enumerate}
    \item The strip of correlations $G_2(x,x')<1$ near the diagonal, $x=x'$, appears due to atomic anti-bunching induced by repulsive interactions \cite{fabbri:wigner,antibunching_paper}. This allows us to verify the correlation sampling by comparing the anti-bunching feature
    obtained with that from an analytical calculations.
    \item The pattern of fringes that run parallel to the diagonal and propagate away from it in time are a result of the interaction quench between $t=0$ ms and $t=2$ ms. In the context of analogue gravity this can be viewed as cosmological particle creation due to the sudden quench \cite{fabbri:wigner,barcelo:cpc,piyush:cpc}.
    \item The two tongues, which emerge from the diagonal at the location $(x,x')=(x_{h},x_{h})$, with $x_h\approx -30$ $\mu$m corresponding to the sonic black hole horizon are the key signature of analogue Hawking radiation in the density-density correlation function \cite{fabbri:wigner,fabbri:corrfct}. These tongues indicate correlation between the two points $x$ and $x'$ on either side of the horizon, due to the presence of the Hawking particle and antiparticle analogues at $x$ and $x'$.
\end{enumerate}
In \frefp{fig:atomic_losses}{b-d}, we have also marked new features of interest 4 and 5 (and changes to 3), through which results including atomic losses qualitatively deviate from the loss-free scenario. These constitute our main results, and are discussed in the subsequent sections.

\subsection{Stronger correlations in the presence of loss} \label{strong_correl_loss}

\begin{figure}[htb]
    \centering
    \includegraphics[width=\columnwidth]{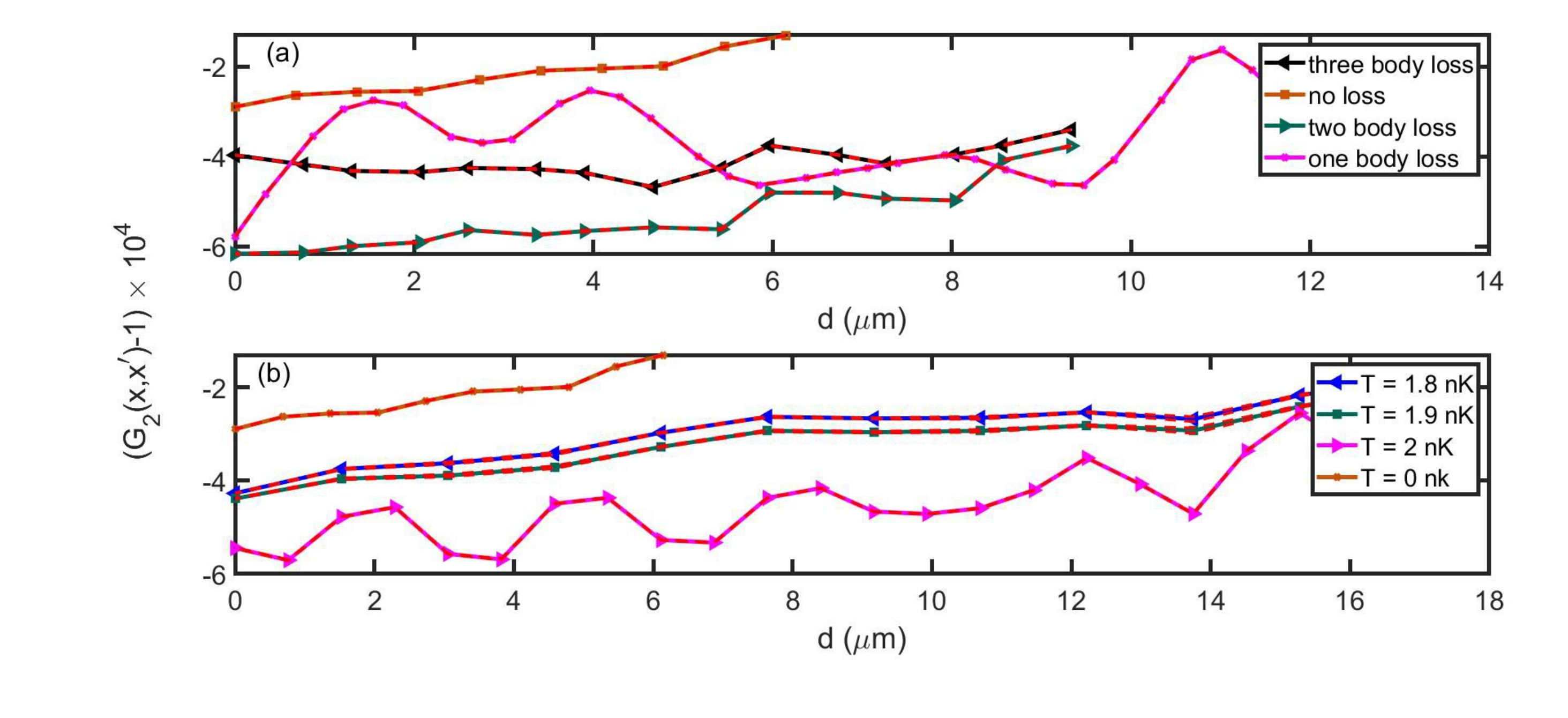}
    \caption{Correlation signal $G_2(x,x')-1$ as a function of the distance $d$ from the diagonal ($x=x'$)
    on a 1D cut along the Hawking tongues, marked as feature 3 and 6 in \fref{fig:atomic_losses}. In (a) we vary the type of loss added as shown in the legend, and in (b) we vary the initial temperature of the condensate, while not including loss. Both panels are for time $t=48$ ms. The sampling error for $\sub{N}{traj}=200000$ trajectories is not visible on the scale of the figure.   
\label{hawking_tongue_cuts}}
\end{figure}
Counter-intuitively, we find that the contrast of the Hawking tongues \emph{increases} with the inclusion of loss in the simulation, pertaining to feature 3 in \frefp{fig:atomic_losses}{d}. To see the effect more clearly, we show one-dimensional cuts of the correlations along the tongue in \frefp{hawking_tongue_cuts}{a}, comparing simulations with all three types of loss.

To understand the physical reason for this, recall that Hawking radiation can also be stimulated \cite{weinfurtner:stimulated_HR}, in the cosmological as well as in analogue systems, instead of being emitted spontaneously \cite{steinhauer:entanglement}. Since one consequence of atom losses is heating of the condensate \cite{dziarmaga:lossheating}, we conjecture that the strengthening of correlations is linked to this heating. This hypothesis is supported by simulations where we compare the cuts along the Hawking tongues for $T\neq 0$ in \frefp{hawking_tongue_cuts}{b} with simulations including loss as in \frefp{hawking_tongue_cuts}{a}. Scenarios starting at finite temperature should also give rise to a larger fraction of stimulated AHR, since in these the Bose-gas contains phonon excitations already from the beginning. We indeed see a similar increase of contrast, as in the lossy scenario, for the temperatures indicated. For example the signal including three body loss in \frefp{hawking_tongue_cuts}{a}, lies in between the results for $T=1.9$ nK and $T=2$ nK, with some deviation in details.

\begin{figure}[htb]
    \centering
    \includegraphics[width=0.99\columnwidth]{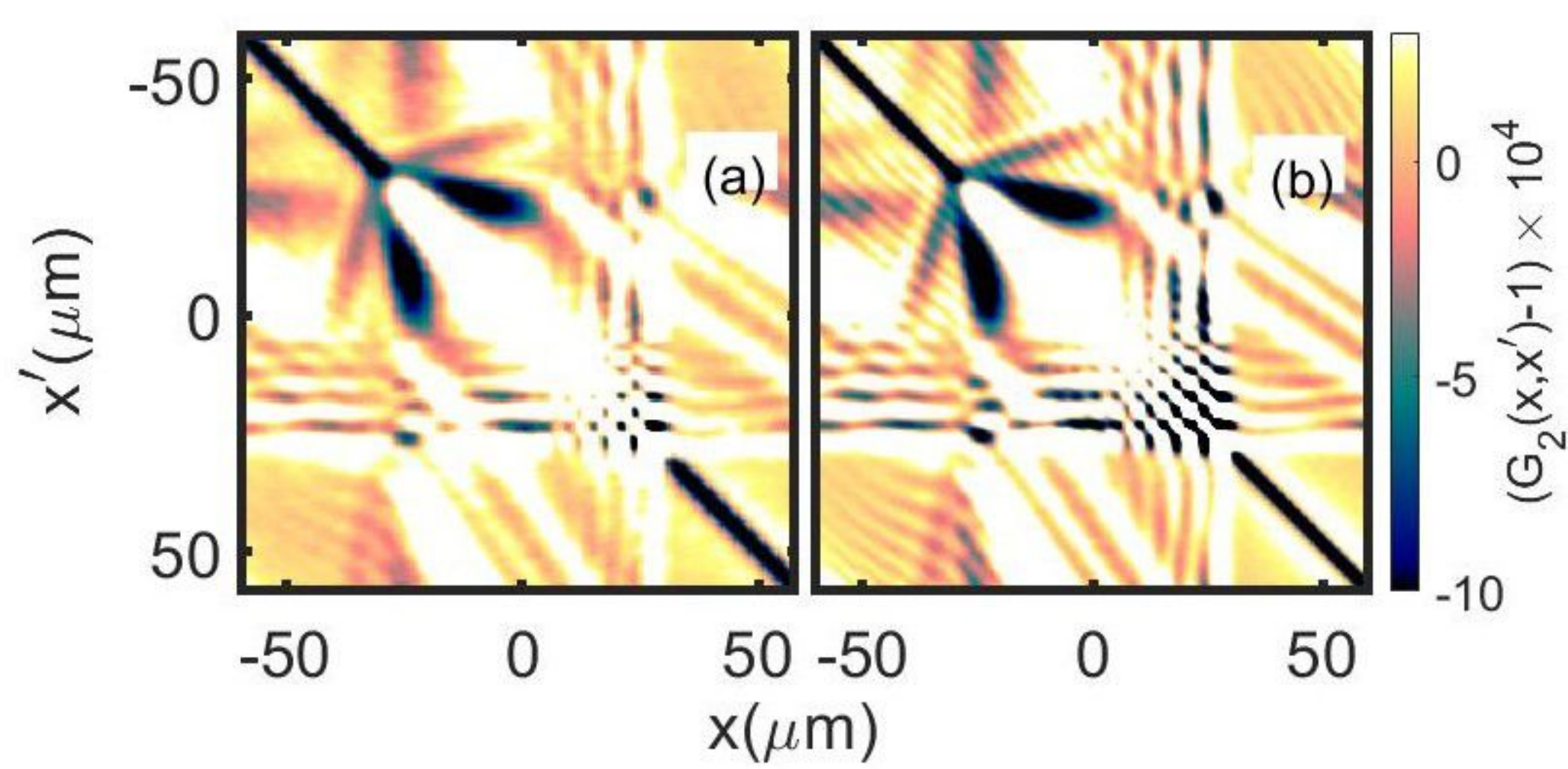}
    \caption{Density-density correlations near a sonic horizon at $x,x'\approx x_h=-30$ $\mu\text{m}$, for non vanishing initial temperatures (a) $T=1.9$ nK and (b) $T=2$ nK at time $t=48$ ms, but not including dynamics loss processes.}
    \label{fig:correlation_with_temperature}
\end{figure}

We also show the entire correlation function for nonzero initial temperatures $T=1.9$ nK and $T=2$ nK, but excluding losses in \fref{fig:correlation_with_temperature}. The closer resemblance of the Hawking tongues including losses in \frefp{fig:atomic_losses}{b-d} with the ones in \fref{fig:correlation_with_temperature} compared to \frefp{fig:atomic_losses}{a} again strengthens the association of signal increase with heating induced losses. Simulations of AHR with a finite initial temperature were also presented in \cite{fabbri:wigner}, demonstrating two tongues, the one due to spontaneous AHR, and a second one due to the reflection of thermal phonons off the horizon.
This is similar to what we observe in \fref{fig:atomic_losses} (b),(c) and (d) at the black hole horizon near $(x_h,x_h)$.

\subsection{Change of slope in presence of loss} \label{slope_change}

Along with an increase in the strength of the Hawking tongues, we notice in \frefp{fig:atomic_losses}{b-d} a change of the slope 
in the $x$,$x'$ plane of the Hawking tongue, marked feature 4. This slope $\alpha$ is dynamically constrained by the propagation velocity of the correlated Hawking phonons in the moving medium that they are immersed in, and is computed as $\alpha=\frac{v_0-c_{\text{sub}}}{v_0-c_{\text{sup}}}= 1$ for the region $x>x'$, in the scenario of \fref{fig:atomic_losses} (a) \cite{fabbri:wigner}. 

We can attribute the variation of the slope to the decrease in the speed of sound in both regions, since loss dynamically reduces the density of the system.
This leads to a decrease in $|v_0-c_{\text{sub}}|$, since $c_{\text{sub}}$ reduces from its original value to become closer to $v_0$, while $|v_0-c_{\text{sup}}|$ increases, as $c_{\text{sup}}$ decreases from its original value to drop further below $v_0$. Hence, $\alpha= \frac{v_0-c_{\text{sub}}}{v_0-c_{\text{sup}}}$ decreases in magnitude, which is what we observe in \fref{fig:atomic_losses} (b), (c) and (d), where the tongues bend inwards towards the diagonal. As an example, for figure \fref{fig:atomic_losses} (d), the mean density has decreased by $36.2$\% when compared to the mean density at $t=0$ ms, decreasing $\sub{c}{sup}$ by a factor of $\approx 1.21$ and $\sub{c}{sub}$ by $\approx 1.19$.

In principle, the variation of the two speeds of sound is not linear in time and hence the Hawking tongue should be curved. However, this curvature is extremely small and hence the Hawking tongue can be well approximated by a line, justifying our use of linear cuts for \fref{hawking_tongue_cuts}. 

\subsection{White hole correlation pattern} \label{whitehole_pattern}

Let us finally discuss feature number 5 in \frefp{fig:atomic_losses}{d}. It is known that the system with a black hole and white hole horizon is dynamically unstable, forming a black hole laser \cite{corley:black_hole_lasers,corley:moddispersion} through the exponential amplification of the superluminal partners of analogue Hawking radiation bouncing back and forth between the horizons. Viewed separately, it is only the white hole that is dynamically unstable 
\cite{mayoral:white_hole,deNova:bhlaser_instability}. The checkerboard pattern visible near the white hole ($x_{w}=3$ $\mu$m) in Figures \ref{fig:atomic_losses} (a) has earlier been attributed to unstable modes of the white hole. 
 
\begin{figure}[htb]
    \centering
    \includegraphics[width=0.9\columnwidth]{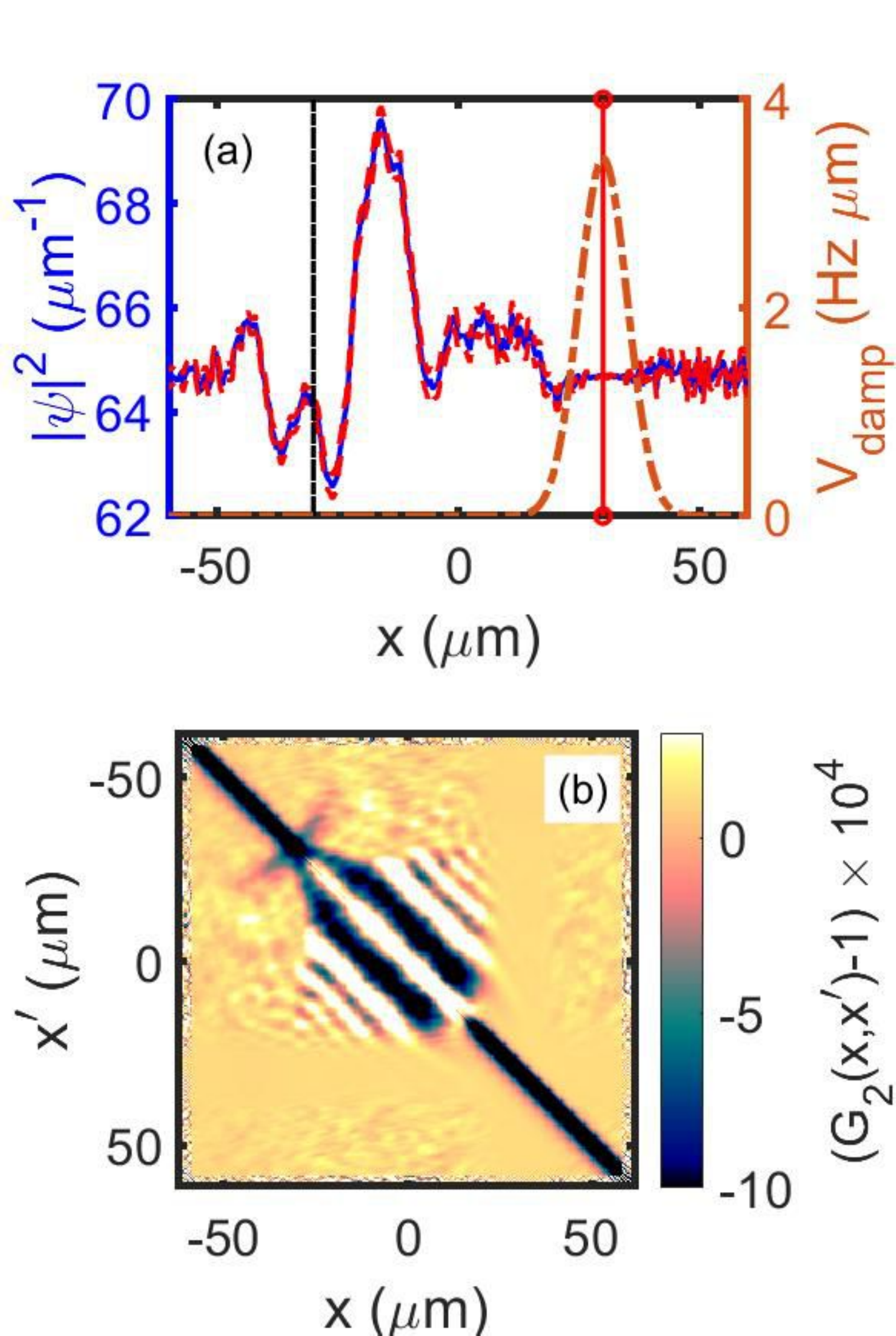}
    \caption{(a) Spatial variation of condensate density (blue) and damping potential (orange), as defined in \eref{damp_pot}. The Black vertical line represents the Black hole horizon while the red represents the White hole horizon. (b) Correlation pattern $G_2(x,x')-1$ in the presence of strong damping at the white hole. The simulation is for a BEC with no losses at $T=0$ nK.}
    \label{fig:WH_damping_plots}
\end{figure}

We see here that atomic losses strengthen the checkerboard pattern, compare \fref{fig:atomic_losses} (a) with \fref{fig:atomic_losses} (d). Our interpretation is again, that this is due to loss induced heating, which creates noise that seeds these instabilities more strongly than the pure vacuum fluctuations in \eref{eq:initial_state}. To demonstrate that the pattern can be attributed to white hole instabilities, we show in \fref{fig:WH_damping_plots}{b} the scenario where strong damping is present at the white hole, which removes the pattern. The density for this scenario is shown in panel (a), together with the damping kernel. Further details about the damping potential can be found in in \aref{app:white_hole_damping}.

%
\section{Conclusions and outlook}  \label{sec:Conclusions_and_outlook}

We have modelled the effect of atom loss in a Bose-Einstein condensate on the correlation signature of analogue Hawking radiation.
For this we used the truncated Wigner approximation to include the dynamics of fluctuations around the mean field.
Counterintuitively, we find that the contrast of the main correlation signal increases due to losses. We attribute this to the additional presence of stimulated Hawking radiation. The latter is an indirect effect, in which the condensate first heats up due to the losses \cite{dziarmaga:lossheating,wuester:phonon_vs_AHR}, and thermally populated fluctuations subsequently stimulate AHR \cite{finke:nonclassical_excitation_in_BECs}. Additional consequences of the same heating effect are a change of slope of the Hawking tongue and a strengthening of the white hole instability pattern.

Our results indicate that measurements of AHR correlations can provide information on additional processes in the Bose-gas, not directly linked to AHR and that spurious stimulated contributions should be taken into account when interpreting experiments. In a next step, it would be interesting to to study the effect of losses on the violation of Cauchy-Schwartz inequalities, which are a tool to flag the spontaneous contribution to AHR as shown in  \cite{de_Nova:Entanglement_and_violation_of_classical_inequalities,Steinhauer:entanglement_proof}. 
\acknowledgments
We gladly acknowledge fruitful discussions with A. Pendse, A. Sreedharan and A. Rana. Financial support from the Max-Planck society under the MPG-IISER partner group program is also gratefully acknowledged.
\appendix
\section{Truncated Wigner treatment of losses} 
\label{TWAlosses}
We now briefly describe the origin of equations \ref{eq:one_body_loss_term}, \ref{eq:two_body_loss_term} and \ref{eq:three_body_loss_term}, with more details available in e.g~Refs.~\cite{ashton:loss,gardiner:Wigner_review}. 

For this purpose we consider the evolution equation due to three body losses \cite{ashton:loss}.
The master equation for the three body recombination process, in the Schr\"{o}dinger picture, is \cite{dziarmaga:lossheating}
\begin{align}
\label{masterequation}
    \pdv{\hat{\rho}}{t} = -\frac{i}{\hbar}\comm{\hat{H}}{\hat{\rho}(t)} + \frac{\gamma_{3,1D}}{6}\int dx[2\hat{\Psi}(x)^3 \hat{\rho} \hat{\Psi}^{\dagger}(x)^3 \CR
    - \hat{\Psi}^{\dagger}(x)^3 \hat{\Psi}(x)^3\hat{\rho} -\hat{\rho}\hat{\Psi}^{\dagger}(x)^3 \hat{\Psi}(x)^3 ]  
\end{align}
where
\begin{align}
    \hat{H}= \int dx \left[ \hat{\Psi}^{\dagger}(x)\left\{-\frac{\hbar^2}{2m}\pdv[2]{x} + V(x) \right\} \hat{\Psi}(x) \right. \CR 
    +  \left. \frac{U_0}{2} \hat{\Psi}^{\dagger}(x) \hat{\Psi}^{\dagger}(x) \hat{\Psi}(x) \hat{\Psi}(x) \right].
\end{align}
We can express the density matrix $\hat{\rho}$ in terms of the Wigner function $W(\psi,\psi^*)$ as \cite{gardiner:Wigner_review} 
\begin{align}
    W(\psi(x),\psi^*(x))\equiv \frac{1}{\pi^2}\int D[\lambda(x)]D[\lambda^*(x)]
    \CR \exp{-\lambda(x)\psi^{*}(x)+
     \lambda^*(x)\psi(x))\chi_{w}(\lambda(x),\lambda^*(x)},    
\end{align}
where $D[\lambda(x)]$ is a functional integration, and the characteristic Function $\chi_w(\lambda(x),\lambda^*(x))$ is given by \cite{gardiner:Wigner_review} 
\begin{equation}
    \chi_{w}(\lambda(x),\lambda^*(x)) = \mbox{Tr}\left[\hat{\rho}\exp{\int dx ( \lambda\hat{\Psi}^{\dagger}(x)-\lambda^{*}\hat{\Psi}(x)) }\right].
\end{equation}
One then converts the equation of motion \bref{masterequation} for the density operator into an equation of motion for the Wigner function. By computing the functional derivatives of the Displacement operator
\begin{equation}
    \hat{D}\equiv \exp{\int dx ( \lambda(x)\hat{\Psi}^{\dagger}(x)-\lambda^{*}(x)\hat{\Psi}(x))    }
\end{equation}
with respect to $\lambda(x)$ and $\lambda^*$, and considering the effect of the same on the equation of motion of the Wigner function, one arrives at the 
functional Wigner operator correspondences \cite{steel:wigner}
\begin{subequations}\label{function_operator_correspondence}
    \begin{align}
       \hat{\Psi}(x) \hat{\rho}  &\rightarrow \left[\psi(x) + \frac{1}{2} \fdv{\psi^*(x)}\right] W(\psi,\psi^*,t),\\
        \hat{\rho} \hat{\Psi}^{\dagger}(x)& \rightarrow \left[\psi^*(x) + \frac{1}{2} \fdv{\psi(x)}\right]W(\psi,\psi^*,t), \\
        \hat{\rho} \hat{\Psi}(x)   &\rightarrow \left[\psi(x) - \frac{1}{2} \fdv{\psi^*(x)}\right] W(\psi,\psi^*,t),\\
        \hat{\Psi}^{\dagger}(x) \hat{\rho} &\rightarrow \left[\psi^*(x) - \frac{1}{2} \fdv{\psi(x)}\right] W(\psi,\psi^*,t).
    \end{align}
\end{subequations}

The resultant equation of motion for $W$ when including losses will contain up to third order partial derivatives with respect to $\psi$ and $\psi^*$, where we discard all down to second order to reach a Fokker-Planck equation (FPE), in the usual truncation scheme:
\begin{align}
    \pdv{W^{TBL}}{t} = \frac{\gamma_{3,1D}}{2}\int dx ( \fdv{\psi}\psi + \fdv{\psi^*_P}\psi^*\CR
     +  3 \fdv{\psi} \fdv{\psi^*} )|\psi|^4 W(\psi(x),\psi^*(x),t).    
\end{align}
with $TBL$ indicating that we consider only terms which arise from \eref{eq:three_body_loss_term}.

Since solutions of a FPE directly correspond to those of a stochastic differential equation (SDE), we can solve the former by expressing it using the SDE
\begin{align}
    d\psi(x) &= \\
    &-\frac{\gamma_{3,1D}}{2} |\psi(x)|^4 \psi \:dt  +\frac{\sqrt{3\gamma_{3,1D}}}{\sqrt{2}} |\psi(x)|^2 
     d\mathscr{W}(x,t).\nonumber
\end{align}
Adding the usual terms unrelated to loss \cite{steel:wigner}, we finally reach
\begin{align}
   & \pdv{\psi}{t} = -\frac{i}{\hbar}\left[ -\frac{\hbar^2}{2m}\pdv[2]{\psi}{x}  +V(x)+ U_0 |\psi|^2 \right] \psi \CR
    &- \frac{\gamma_{3,1D}}{2} |\psi|^4 \psi \:dt + \frac{\sqrt{3\gamma_{3,1D}}}{\sqrt{2}} |\psi|^2 d\mathscr{W}.
\end{align} 
Similar derivations for one- and two-body loss processes yield \eref{eq:one_body_loss_term} and \eref{eq:two_body_loss_term}.

\section{Truncated Wigner treatment of correlations} 
\label{TWA_correlations}
As stated before, the TWA allows the sampling of quantum correlations through symmetrically ordered stochastic averages \cite{ashton:loss,gardiner:Wigner_review}.
In this appendix we describe how these can be assembled to infer the correlation function \bref{eq:G_2_correlation} 
that is central to the present work. The starting point is the association
\begin{equation}
    \overline{ \psi^*(x) \psi(x') } = \frac{1}{2}\langle\hat{\Psi}^{\dagger}(x) \hat{\Psi}(x') + \hat{\Psi}(x')\hat{\Psi}^{\dagger}(x) \rangle, 
\end{equation}
where the dependence on time has been suppressed since we will deal with equal time correlations only. With the commutation relation $\delta_p(x,x') = \comm{\Psi(x)}{\Psi^{\dagger}(x')}$, we obtain
\begin{equation}\label{eq:G_1_correlation}
     \langle\hat{\Psi}^{\dagger}(x) \hat{\Psi}(x')\rangle = \langle \psi^*(x) \psi(x') \rangle_W - \frac{1}{2}\delta_p(x,x'),
\end{equation}
providing already first order phase correlations $G_1(x,x')=  \langle\hat{\Psi}^{\dagger}(x) \hat{\Psi}(x')\rangle$. Here $\delta_p(x,x')$ is a restricted basis delta function given by \cite{gardiner:Wigner_review}
\begin{equation}
    \delta_p(x,x') = \frac{1}{L}\sum_{k} \left[ u^2_k e^{ik(x-x')} - v^2_k e^{-ik(x-x')}\right],
\end{equation}
where the index $k$ enumerates the finite number of Bogoliubov modes onto which we add noise for the numerical simulation, in \eref{eq:initial_state}. The expression converges to the actual delta function for $k\rightarrow \infty$. 

In a similar fashion, we can relate $ \langle\hat{\Psi}^{\dagger}(x) \hat{\Psi}^{\dagger}(x') \hat{\Psi}(x) \hat{\Psi}(x')\rangle$ with $\overline{ \psi^*(x)\psi^*(x')\psi(x)\psi(x') }$. We first write the latter as a symmetric sum of 24 averages containing all the possible permutations of field operators. Each can be brought into the form $\langle\hat{\Psi}^{\dagger}(x) \hat{\Psi}^{\dagger}(x') \hat{\Psi}(x) \hat{\Psi}(x')\rangle$ using the commutation relation. After some algebra, we finally obtain 
\begin{align}
   & \langle\hat{\Psi}^{\dagger}(x) \hat{\Psi}^{\dagger}(x') \hat{\Psi}(x) \hat{\Psi}(x')\rangle = \overline{ \psi^*(x)\psi^*(x')\psi(x)\psi(x') } \CR
   &- \frac{1}{2}[\delta_p(x,x') \overline{ \psi^*(x) \psi(x')}  + \delta_p(x',x') \overline{ \psi^*(x) \psi(x)}  \CR
   &+  \delta_p(x,x) \overline{ \psi^*(x') \psi(x')}  + \delta_p(x',x) \overline{ \psi^*(x') \psi(x)}  ] \CR
   &+ \frac{1}{4}\left[\delta_p(x,x) \delta_p(x',x')+ \delta_p(x,x') \delta_p(x',x)\right].
\end{align}
%

\section{White hole damping}\label{app:white_hole_damping}
\label{WH_damping} 
In this appendix, we describe our implementation of damping on the white hole. For this we add a complex potential 
\begin{align} 
\sub{V}{damp}(x) &=  -i \frac{s}{\hbar}\exp \left (-\frac{{(x-x_{w})}^2}{2\sigma^2_d} \right) (|\psi(x)|^2-\rho_0)\psi
\label{damp_pot}
\end{align}
to the right hand side of \eref{eq:GPE}. Here $x_{w}$ is the location of the white hole horizon, $s$ the damping strength and $\sigma_d$ the width of the damping profile while $\rho_0$ is as defined in \eref{eq:initial_state}. 

One can see, that \bref{damp_pot} causes exponential damping of $\psi$ if the local density at the white hole deviates from the mean value $\rho_0$.
Since such deviations are integral to unstable modes, the growth of the latter is damped.

\section{Dimensionality reduction}\label{app:dimensionality_reduction_appendix}
Here we briefly discuss the reduction of the 3D equation of motion to an effective 1D equation. For this purpose, we rewrite the fieldoperator
\begin{align}
    \hat{\Psi}(x,y,z) =  \frac{1}{\sqrt{\pi \sigma_y \sigma_z}} e^{-\frac{y^2}{2\sigma^2_y}} e^{-\frac{z^2}{2\sigma^2_z}} \hat{\Psi}(x),
\end{align}
such that transverse excitations are frozen out, using $\sigma_y=\sqrt{\hbar/(m\omega_y)}$, $\sigma_z=\sqrt{\hbar/(m\omega_z)}$, with $\omega_y$ and $\omega_z$ the trapping frequencies in the $y$ and $z$ directions, respectively. Defining $\mathscr{N} = \frac{1}{\sqrt{\pi \sigma_y \sigma_z}} $, we obtain that e.g.~
\begin{align}
        &\int_{-\infty}^{\infty} \int_{-\infty}^{\infty} \int_{-\infty}^{\infty} dxdydz \hat{\Psi}^{\dagger}(x,y,z) \hat{\rho} \hat{\Psi}(x,y,z) = \\  &{\mathscr{N}}^2\pi\sigma_y\sigma_z \int_{-\infty}^{\infty} dx \hat{\Psi}^{\dagger}(x) \hat{\rho} \hat{\Psi}(x) \nonumber,
\end{align}
Thus, the 1D master equation for one body loss is
\begin{equation}
     \pdv{\hat{\rho}}{t} = \gamma_{1,3D}  \int dx[2\hat{\Psi}(x) \hat{\rho} {\hat{\Psi}^{\dagger}(x)} - {\hat{\Psi}^{\dagger}(x)} \hat{\Psi}(x)\hat{\rho} -\hat{\rho}{\hat{\Psi}^{\dagger}(x)} \hat{\Psi}(x) ].
\end{equation}
Similarly we reach
\begin{align}
    \pdv{\hat{\rho}}{t} = & \gamma_{2,3D}  {\mathscr{N}}^4\frac{\pi\sigma_y\sigma_z}{2} \int dx[2\hat{\Psi}(x)^2 \hat{\rho} \hat{\Psi}^{\dagger}(x)^2 \\
    &- \hat{\Psi}^{\dagger}(x)^2 \hat{\Psi}(x)^2\hat{\rho} -\hat{\rho}\hat{\Psi}^{\dagger}(x)^2 \hat{\Psi}(x)^2 ] \nonumber,
\end{align}
for two-body loss and
\begin{align}
    \pdv{\hat{\rho}}{t} = &\frac{\gamma_{3,3D}}{6} {\mathscr{N}}^6 \frac{(\pi\sigma_y\sigma_z)}{3} \int dx[2\hat{\Psi}(x)^3 \hat{\rho} \hat{\Psi}^{\dagger}(x)^3 \\ 
    & -\hat{\Psi}^{\dagger}(x)^3 \hat{\Psi}(x)^3\hat{\rho} -\hat{\rho}\hat{\Psi}^{\dagger}(x)^3 \hat{\Psi}(x)^3 ] \nonumber   
\end{align}
for three-body loss. At this point we can define effective 1D loss rates
\begin{align}\label{eq:reduced_loss_rates}
    &\gamma_{1,1D} = \gamma_{1,3D},\\
    &\gamma_{2,1D} = \gamma_{2,3D}  \frac{{\mathscr{N}}^4\pi\sigma_y\sigma_z}{2} = \frac{\gamma_{2,3D}}{2{(\pi\sigma_y\sigma_z)}}, \\
    &\gamma_{3,1D} = \gamma_{3,3D} \frac{{\mathscr{N}}^6\pi\sigma_y\sigma_z}{3} = \frac{\gamma_{3,3D}}{3{(\pi\sigma_y\sigma_z)}^2},
\end{align}
which are used in the main article.

\bibliography{losshawking}
\end{document}